\documentclass[journal=nalefd,manuscript=letter]{achemso}

\usepackage[version=3]{mhchem}
\usepackage{caption}
\usepackage{float}
\usepackage{geometry}
\usepackage{setspace}
\usepackage{array}
\usepackage{multirow}
\usepackage{adjustbox}
\usepackage{longtable}
\usepackage{graphicx}
\usepackage{epstopdf}
\usepackage{subcaption}
\usepackage[usenames,dvipsnames]{xcolor}
\usepackage{comment}

\usepackage{makecell}

\newcolumntype{?}{!{\vrule width 1.2pt}}

\author{Shashikant Kumar}
\author{Phanish Suryanarayana}
\email{phanish.suryanarayana@ce.gatech.edu}
\affiliation[Georgia Tech]{College of Engineering, Georgia Institute of Technology, Atlanta, GA 30332, USA}

\title{Bending moduli for thirty-two select atomic monolayers from first principles}

\begin{document}
	
	
\begin{abstract}
We calculate bending moduli along the principal directions for thirty-two select atomic monolayers using ab initio Density Functional Theory (DFT). Specifically, considering representative  materials from each of Groups IV, V, III-V monolayers,   transition metal dichalcogenides, Group III monochalcogenides, Group IV monochalcogenides, and transition metal trichalcogenides, we utilize the recently developed Cyclic DFT method to calculate the bending moduli in the practically relevant but previously intractable low-curvature limit. We find that the moduli generally increase with thickness of the monolayer and that structures with a rectangular lattice are prone to a higher degree of anisotropy relative to those with a honeycomb lattice. We also find that exceptions to these trends are generally a consequence of unusually strong/weak bonding  and/or significant structural relxation related effects. 
\end{abstract}

\newpage

The experimental discovery of graphene nearly two decades ago \cite{Novoselov10451} has had a  revolutionary effect on the field of two-dimensional materials, with dozens of atomic monolayers having now been synthesized \cite{balendhran2015elemental, zhou2018monolayer, vaughn2010single,zhang2018recent,dai2016group} and the potential for thousands more as predicted by ab initio calculations \cite{haastrup2018computational,zhou20192dmatpedia}. These materials have been the subject of intensive research \cite{mas20112d,butler2013progress,geng2018recent},  inspired by their exciting and fascinating properties that are typically muted or non-existent in traditional bulk counterparts, including unprecedented mechanical strength \cite{balendhran2015elemental}, rare p-type electronic behavior \cite{zhou2018monolayer}, negative Poisson ratio \cite{zhang2018recent}, large piezoelectric effect \cite{fei2015giant}, topological superconductivity \cite{hsu2017topological},  high electron mobility in combination with direct band gap \cite{dai2016group}, and high chemical as well as thermal stability \cite{jiang2015recent}.  Such properties make atomic monolayers ideally suited for a number of technological applications, including flexible electronics \cite{pu2012highly, lee2013flexible, salvatore2013fabrication, yoon2013highly}, nanoelectromechanical devices \cite{zhang2015single, sakhaee2008potential, sazonova2004tunable, bunch2007electromechanical}, and nanocomposites \cite{novoselov2012two,qin2015lightweight}, wherein behavior under   bending deformations is particularly important \cite{lindahl2012determination}.  However, accurate estimates of a property as fundamental as even the bending modulus is not well established for these materials, providing the motivation for the current work. 

Experimental data for the bending moduli of atomic monolayers  is extremely sparse, likely due to the challenges associated with the experimental setup as well as the high accuracy required in measurements \cite{akinwande2017review}. In fact, to the best of our knowledge, only the moduli of graphene \cite{han2019ultrasoft,lindahl2012determination,nicklow1972lattice} have been reported to date, that too with significant error bars. In terms of computations, widely used ab initio methods like Density Functional Theory (DFT) \cite{Hohenberg,Kohn1965} provide an avenue for the accurate calculation of such material properties \cite{kudin2001c,zhao2015two,lai2016bending,nepal2019first}. However, the large computational cost associated with DFT and its cubic scaling with system size restricts the calculations to curvatures that are significantly larger than those  typically encountered in practice \cite{lindahl2012determination,ghosh2019symmetry}. A more efficient but less accurate alternative is to employ atomistic force field simulations  \cite{arroyo2004finite,koskinen2010approximate,cranford2009meso, cranford2011twisted,roman2014mechanical,liu2011interlayer,xu2010geometry,sajadi2018size,qian2020multilayer},. However, the resolution they provide is insufficient for the study of nanoscale systems, particularly in the presence of complex bonding. This is evident by the significant scatter in bending moduli  for even elemental monolayers, with values for graphene ranging from 0.8 to 2.7 eV \cite{arroyo2004finite,sajadi2018size} and for silicene from 0.4 to 38 eV \cite{roman2014mechanical,qian2020multilayer}. 

\begin{figure}[h!]
\includegraphics[scale=0.82]{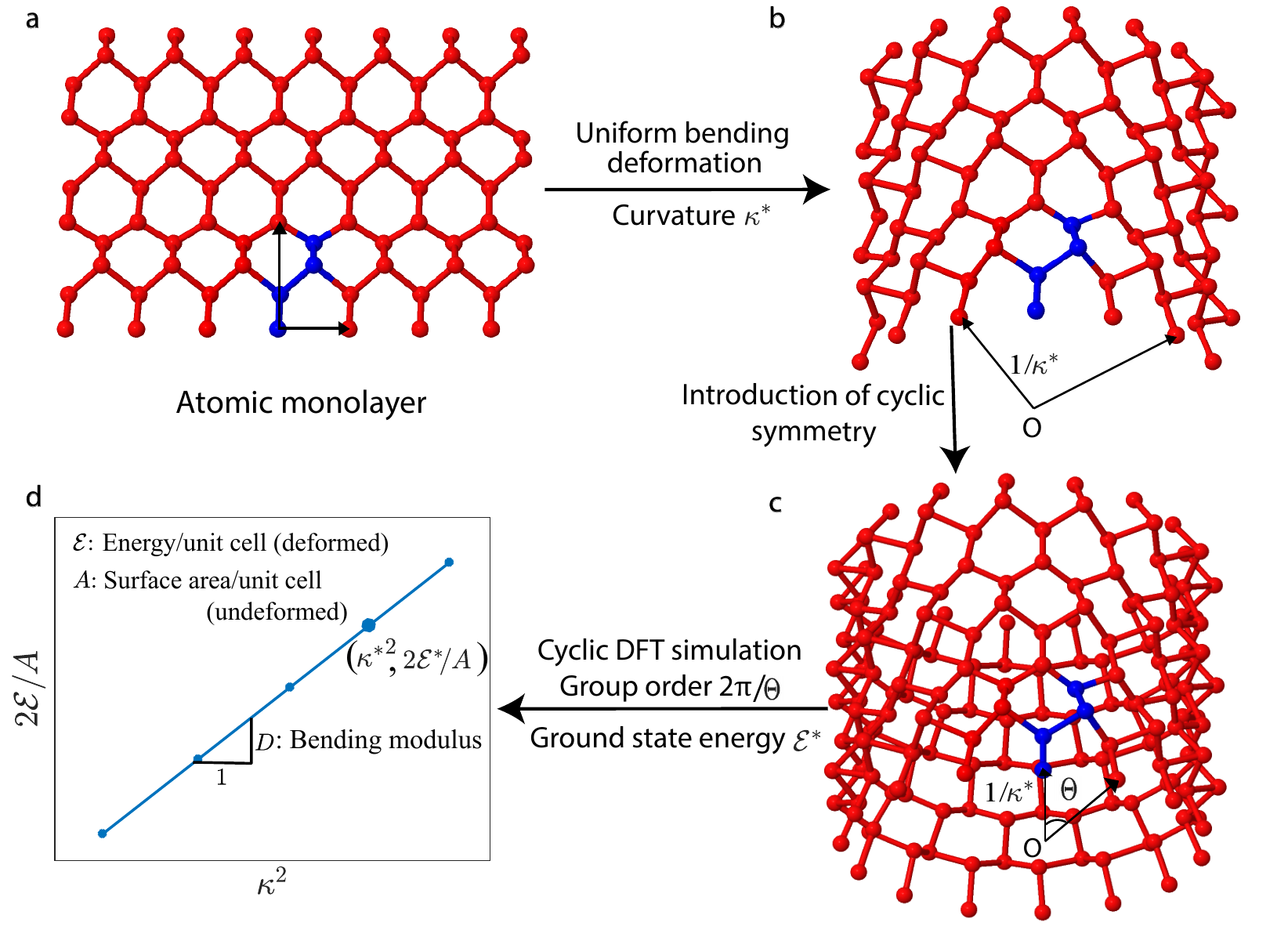}
\caption{Schematic illustrating the calculation of bending moduli for atomic monolayers using the ab initio Cyclic DFT method. The atoms in the unit cell are colored blue. }
\label{fig:illustration}
\end{figure}

Cyclic DFT \cite{banerjee2016cyclic,ghosh2019symmetry} is a recent ab initio method that tremendously reduces the simulation cost for systems with cyclic symmetry. In particular, by utilizing the connection between uniform bending deformations and cyclic symmetry, the method enables the calculation of bending moduli for nanostructures in the practically relevant but previously intractable low-curvature limit. In this approach,  shown schematically in Fig.~\ref{fig:illustration}, a uniform bending deformation  is applied to the nanostructure's unit cell, which is then mapped periodically in the angular direction. The cyclic symmetry of the resultant structure is then exploited to perform highly efficient DFT calculations for the ground state energy. Having done so for different deformations, the bending modulus is determined from the energy's dependence on curvature.  Note that this approach neglects edge-related effects  in the  bending direction,  justified by the nearsightedness of matter \cite{prodan2005nearsightedness} and Saint-Venant's principle \cite{iesan2006saint} at the electronic and continuum scales, respectively. Such assumptions are inherent to ab initio calculations for bulk properties, e.g., surface effects are neglected in Young's modulus calculations.\cite{PhysRevB.87.035423}

In this work, we use the first principles Cyclic DFT method as implemented in the state of the art code SPARC-X---most recent and highly optimized  version of the real-space DFT code SPARC \cite{xu2020sparc,ghosh2017sparc1,ghosh2017sparc2}---to calculate bending moduli along the principal directions for thirty-two select atomic monolayers. Specifically, we consider representative materials with honeycomb lattice structure from each of Groups IV, V, III-V monolayers, transition metal dichalcogenides (TMDs), and Group III monochalcogenides, as well as materials with rectangular lattice structure from each of Group V monolayers, Group IV monochalcogenides, and transition metal trichalcogenides (TMTs). These groups have been selected because  of the significant success in the synthesis of affiliated monolayers, which are found also to demonstrate exotic and novel properties \cite{dai2016group,Coleman568,Novoselov10451,zhou2018monolayer,vaughn2010single,zhang2018recent,balendhran2015elemental}.  Note that the accuracy of the Cyclic DFT formulation and implementation has been systematically and thoroughly benchmarked against established planewave \cite{ABINIT} and real-space \cite{ghosh2017sparc1,ghosh2017sparc2} codes, with representative results recently published in literature \cite{ghosh2019symmetry}, substantiating the validity of the computations performed here. 

In all simulations, we employ the PBE \cite{perdew1986accurate} variant of the GGA   exchange-correlation functional and ONCV \cite{hamann2013optimized} pseudopotentials from the SG15 \cite{SCHLIPF201536} collection. All numerical parameters including grid spacing, number of points for Brillouin zone integration, vacuum in the radial direction, and structural relaxation tolerances (both cell and atom) are chosen such that the computed bending moduli are accurate to within 1\%. In order to arrive at values for mildly bent sheets, i.e., corresponding to the low curvature limit, we choose curvatures in the range $0.12 \leq \kappa \leq 0.24$ nm$^{-1}$, commensurate with those found in experiments\cite{lindahl2012determination}.  At these curvatures, the desired precision in bending modulus translates to the ground state energy being converged to within $10^{-5}$ Ha/atom, necessary to capture the extremely small energy differences.  Note that these highly converged large-scale DFT calculations are prohibitively expensive even with state of the art codes \cite{ghosh2017sparc2,banerjee2018two,motamarri2020dft} on the largest supercomputers. For example, the ZrTe$_3$ system with $\kappa=0.16$ nm$^{-1}$ has a total of $12,120$ electrons and 25 $\mathbf{k}$-points, making it intractable to traditional DFT implementations.

\begin{figure}[H]
	    \centering
		\includegraphics[width=\textwidth]{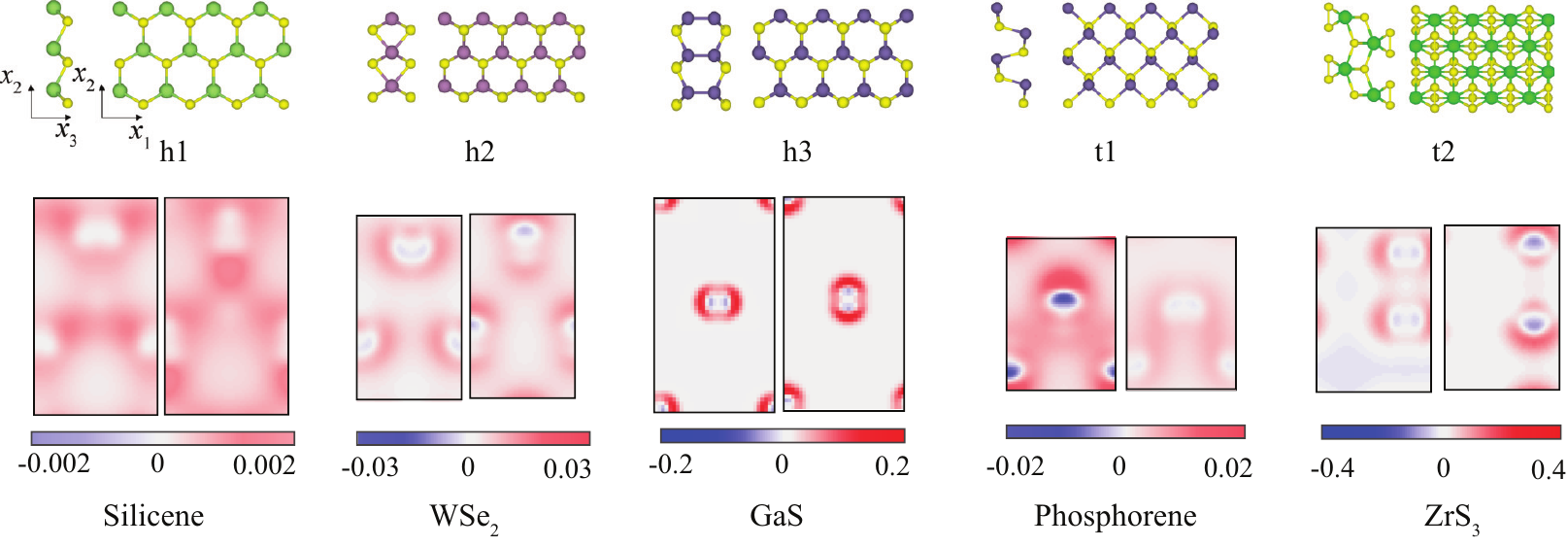}
		\caption{Contours of electron density difference between the flat and bent  ($\kappa \sim 0.2$ nm$^{-1}$) atomic monolayers. The contours are in the undeformed configuration and on the $x_1 x_2$-plane passing through the furthest atom from the monolayer center, with the figures to the left and right for each material corresponding to bending along the $x_1$ and $x_2$ directions, respectively.}
		\label{fig:struc:cont}
\end{figure}

We first determine the suitability of the chosen exchange-correlation functional and pseudopotentials for the thirty-two atomic monolayers being studied here. Specifically, starting from the structures mentioned in Table~\ref{Table:BM}, which are illustrated in Fig~\ref{fig:struc:cont}, we calculate the equilibrium geometry for the monolayers using the planewave DFT code ABINIT \cite{ABINIT}. The relaxed structures so obtained, details of which are presented in the Supporting Information, are found to be in  very good agreement with previous theoretical predictions as well as experimental measurements \cite{dai2016group,Coleman568,Novoselov10451,zhou2018monolayer,vaughn2010single,zhang2018recent,balendhran2015elemental,zhou20192dmatpedia}. Having thus verified the fidelity of the DFT simulations,  we now calculate the bending moduli for the monolayers along their principal directions using the aforedescribed Cyclic DFT methodology. We present the results so obtained in Table~\ref{Table:BM}, where $D_1$ and $D_2$ are used to denote the bending moduli along the principal directions $x_1$ and $x_2$, respectively. The orientation of these directions relative to the  different structures can be seen in Fig.~\ref{fig:struc:cont}. Indeed, for materials with a honeycomb lattice, the  $x_1$ and $x_2$ directions correspond to the zigzag and armchair directions, respectively.  Note that the bending moduli have been normalized with the surface area of the flat monolayer, as is typical while reporting values for two-dimensional materials.\vspace{3mm}

\begin{table}[h!]
    \centering
    \resizebox{1.0\textwidth}{!}{
    \begin{tabular}{?c|c|c|c?c|c|c|c?}
   \Xhline{1.2pt}
   \multirow{3}{*}{Group} &\multirow{3}{*}{Material}&\multicolumn{2}{c?}{Bending}&\multirow{3}{*}{Group}&\multirow{3}{*}{Material}&\multicolumn{2}{c?}{Bending}\\
    &&\multicolumn{2}{c?}{moduli (eV)}&&&\multicolumn{2}{c?}{moduli (eV)}\\
 
    & &\multicolumn{1}{c}{ $D_1$} &\multicolumn{1}{c?}{$D_2$} &&  & \multicolumn{1}{c}{$D_1$} & \multicolumn{1}{c?}{$D_2$}\\
    \Xhline{1.2pt}
    \multirow{2}{*}{ Groups IV, V, III-V}& Si& 0.36 & 0.37 & \multirow{2}{*}{Group III}&GaTe & 14.9 & 14.5 \\
    \cline{2-4}
    \cline{6-8}
    \multirow{2}{*}{monolayers}& P & 0.71 & 0.72& \multirow{2}{*}{monochalcogenides}  & GaSe & 24.1& 18.9  \\
    \cline{2-4}
    \cline{6-8}
    \multirow{2}{*}{(h1)}& BN & 0.56 & 0.58 & \multirow{2}{*}{(h3)}& GaS & 21.2 & 21.7\\
    \cline{2-4}
    \cline{6-8}
    & C & 1.51 & 1.50 &  & InSe& 17.5 & 15.3 \\
    \hline
     & ZrS$_2$ &  0.62 & 0.80 & \multirow{2}{*}{Group V} &  Bi & 3.55& 1.05 \\
    \cline{2-4}
    \cline{6-8}
    \multirow{8}{*}{Transition metal}& ZrSe$_2$ &  2.07 & 2.10 &  \multirow{2}{*}{monolayers} & Sb & 3.95 & 1.14 \\
    \cline{2-4}
    \cline{6-8}
    \multirow{8}{*}{dichalcogenides} &ZrTe$_2$ & 2.84  & 2.16 & \multirow{2}{*}{(t1)}  & As & 5.70& 1.44 \\
    \cline{2-4}
    \cline{6-8}
    \multirow{8}{*}{(h2)}&  TiS$_2$ & 2.62 & 2.47 &  & P & 7.59& 1.44\\
    \cline{2-8}
    & TiSe$_2$ & 2.94 &  2.92 & \multirow{2}{*}{Group IV}& GeS & 3.57& 1.21 \\
    \cline{2-4}
    \cline{6-8}
     & TiTe$_2$ & 4.45 & 4.83 &\multirow{2}{*}{monochalcogenides} & GeSe & 3.91& 1.35\\
    \cline{2-4}
    \cline{6-8}
    & MoS$_2$ &9.12 & 8.98  & \multirow{2}{*}{(t1)} & SnS & 4.26& 2.27 \\
    \cline{2-4}
    \cline{6-8}
    & MoSe$_2$&  9.07&  9.21& & SnSe & 4.23 & 2.24 \\
    \cline{2-8}
   & WS$_2$ & 10.3 & 10.1 & \multirow{2}{*}{Transition metal} & ZrS$_3$ & 23.0& 26.5 \\
    \cline{2-4}
    \cline{6-8}
     & WSe$_2$& 11.4 & 11.1& \multirow{2}{*}{trichalcogenides} & TiS$_3$ & 30.4& 28.5\\
    \cline{2-4}
    \cline{6-8}
     & MoTe$_2$ & 11.4& 11.0 &\multirow{2}{*}{(t2)} & ZrSe$_3$ & 33.6 & 27.5 \\
    \cline{2-4}
    \cline{6-8}
     & WTe$_2$ & 12.8&  12.7&  & ZrTe$_3$ & 24.5 & 76.2\\
     \Xhline{1.2pt}
    \end{tabular}
    }
    \caption{Bending moduli along principal directions for the thirty-two select atomic monolayers from first principles.}
    \label{Table:BM}
    \end{table}

The bending moduli presented here are in reasonably good agreement with previous such ab initio DFT studies. Specifically, the bending modulus of graphene has previously been reported to be $1.46$ eV \cite{kudin2001c}, which is in good agreement with the value of $\sim 1.5$ eV here.\footnote{The computed bending modulus for graphene is also in good agreement with recent experimental measurements \cite{han2019ultrasoft}, where the value is reported to be in the range 1.2 to 1.7 eV. } The same study has reported the bending modulus of BN to be $1.29$ eV \cite{kudin2001c}, which differs from the value here by $\sim 0.72$ eV. In the case of TMDs, the bending moduli values reported recently \cite{lai2016bending} differ by a maximum of $\sim 1.07$ eV from those here, while  having very good  qualitative agreement in the variations between the different monolayers. The quantitative differences between current work and literature can be attributed to the combined effect of a number of factors, including the accuracy of the calculations, choice of bending curvatures, and the nature of structural relaxation performed.  Note that though there are a few other works which calculate the bending moduli of atomic monolayers using DFT \cite{zhao2015two,ghosh2019symmetry,nepal2019first}, here we have only explicitly compared against those that employ the same exchange-correlation functional, i.e., PBE, since values can vary based on this choice.

We observe from the results that the bending moduli span nearly three orders of magnitude between the different monolayers, silicene being at one end of the spectrum with $\sim 0.36$ eV  and ZrTe$_3$ at the other end with $\sim 76.2$ eV. Graphene is towards the lower end, being only a few fold larger than silicene and more than an order of magnitude smaller than ZrTe$_3$. In terms of groups,  Group III monochalcodenides and TMTs have the largest bending moduli among structures with honeycomb and rectangular lattices, respectively. Specifically, the values for both these groups are similar, with TMTs having larger moduli overall. We also observe  from the results that, apart from TMTs, groups with a rectangular lattice have a higher degree of anisotropy compared to those with a honeycomb lattice. Specifically, while the bending moduli of honeycomb structures are generally similar in the armchair and zigzag directions,  rectangular structures have up to  a five-fold difference in the values along the principal directions. This trend is epitomized by phosphorene, which demonstrates significant anisotropy in the rectangular but not honeycomb lattice configurations.

The above observations regarding the variation in bending moduli between the different monolayers as well as the  different directions within monolayers merit further discussion. Since the exact values are determined by the complex interplay between structure, composition, and electronic structure effects, here we focus on general trends. If the monolayers are modeled as plates in the spirit of traditional continuum mechanics \cite{reddy2006theory}, then the bending moduli  are expected to scale cubically with the thickness. Though the exact value of thickness for monolayers like graphene is controversial \cite{huang2006thickness}, in the current context we correlate it with the spatial extent of the electron density in DFT. However, such a continuum formulation neglects the highly inhomogeneous nature of the bonding that can occur in atomic monolayers, rendering the scaling laws unrepresentative for these systems, a conclusion that is indeed supported by the data here. It is nevertheless true that in the absence of structural relaxation, there is larger distortion in the bonding along the bending direction as one moves away from the center of the monolayer (i.e., neutral axis). This translates to increase  in bending moduli with monolayer thickness,  exceptions being materials that have unusually strong (e.g. graphene and BN \cite{demczyk2002direct,wang2017graphene}) or weak (e.g. TiS$_2$ and TiTe$_2$) bonding, and/or those with significant structural relaxation related effects (e.g., TMTs, as discussed below).\footnote{The bending moduli values in the absence of structural relaxation can be found in the Supporting Information.} 


In view of the above discussion, we plot in Fig.~\ref{fig:struc:cont}  contours of electron density difference between the flat and bent monolayers for silicene, WSe$_2$, GaTe, phosphorene, and ZrS$_3$: materials with  varied structure that have bending moduli spanning the range of values  reported here. The contours are plotted in the undeformed configuration and on the $x_1 x_2$-plane passing through the furthest atom from the monolayer center. It is clear that the electron density perturbations generally increase with monolayer thickness,  resulting in the bending moduli following a similar trend. In addition, there is noticeable difference in the electron density perturbations between the two bending directions for phosphorene, leading to its significant anisotropy.  Similar to previous conclusions  for various properties \cite{li2019emerging, xia2014rediscovering, wang2015highly, luo2015anisotropic}, this can be attributed to the lower degree of structural symmetry in rectangular lattices. Note that TMTs are exceptions to these trends,   as evident from the contours for ZrS$_3$, wherein  the electron density perturbations are similar for both bending directions. We have found that this is a consequence of structural relaxation related effects: there is a drastic drop in the value of $D_2$ for TMTs, e.g.,  the value for TiS$_3$ drops from 87.4 to 28.5 eV and the value for ZrS$_3$ drops from 82.6 to 26.5 eV. In particular, we have found that  the  distance between the  two atoms furthest from the neutral axis is significantly reduced after relaxation, taking a value  similar to that obtained for bending along the $x_1$ direction, both of which are close to the distance in the flat sheet. This essentially negates the effect of the bending  deformation in the region where it is expected to have the maximum effect, thereby resulting in the drastic reductions. Note that the significant anisotropy observed for honeycomb structured  ZrS$_2$ and GaSe can similarly be explained in terms of structural relaxation related effects, as evident from the unrelaxed bending moduli values presented in the Supporting Information.

In summary, we have calculated bending moduli along the principal directions for thirty-two select atomic monolayers from first principles. In particular, we have used the recent Cyclic DFT method to calculate the bending moduli of the two-dimensional materials in the practically relevant but previously intractable low-curvature limit. We have found that the moduli generally increase with monolayer thickness, spanning nearly three orders of magnitude between the different materials. In addition, we have found that monolayers with rectangular lattice structures are prone to a higher degree of anisotropy relative to those with honeycomb lattices. Exceptions to these trends generally result from unusually strong/weak bonding  and/or significant structural relxation related effects. Overall, this work provides an important reference for the bending moduli of a number of important  atomic monolayers.

\section*{Supporting Information}
Equilibrium geometries of atomic monolayers, unrelaxed bending moduli

\section*{Acknowledgments}
The authors gratefully acknowledge the support of the U.S. National Science Foundation (CAREER--1553212).

\providecommand{\latin}[1]{#1}
\makeatletter
\providecommand{\doi}
  {\begingroup\let\do\@makeother\dospecials
  \catcode`\{=1 \catcode`\}=2 \doi@aux}
\providecommand{\doi@aux}[1]{\endgroup\texttt{#1}}
\makeatother
\providecommand*\mcitethebibliography{\thebibliography}
\csname @ifundefined\endcsname{endmcitethebibliography}
  {\let\endmcitethebibliography\endthebibliography}{}


\end{document}